\documentstyle[prl,aps,twocolumn,floats]{revtex}
\begin{document}
\input epsf
\draft
\title{ESS and Dissipation Range Dynamics of 3-D Turbulence}

\author{ Anirban Sain$^1$  and J.K. Bhattacharjee$^2$} 
\address{$^1$Department of Physics, Indian Institute of Science,
Bangalore - 560 012, India}
\address{$^2$Department of Theoretical Physics, Indian Association
for Cultivation of Science,Jadavpur, Calcutta - 700 032, India} 
\date{October 28th,1997}
 
\maketitle

\begin{abstract}
We carry out a self consistent calculation of the structure 
functions in the dissipation range using Navier Stokes equation.
Combining these results with the known structures in the inertial
range, we actually propose crossover functions for the structure 
functions that takes one smoothly from the inertial to the 
dissipation regime. In the process the success of the extended 
self similarity is explicitly demonstrated.
\end{abstract}


The inertial range of fully developed homogeneous, isotropic 
turbulece has been investigated extensively [1-8]
in the past decade. In comparison far dissiaption range is been
less well studied and as far as we know, a systematic study of
the structure functions, based on Navier Stokes equation (NS),
has not been carried out. In this work, we report a self consistent 
calculation of the structure functions in the dissipation range.
Using accepted results in the inertial range, we propose crossover 
functions for the structure functions and thus demonstrate how
extended self similarity can be understood.\\

We work with forced three dimensional NS equation for 
incompressible flows, written in the momentum space as,

\begin{equation}
\dot{v}_{i}({\bf k}) + \nu k^{2}v_i ({\bf k}) =
\frac{-i}{2}M_{ijl} ({\bf k})\sum_{\bf p}v_{j}({\bf p})
v_{l}({\bf k-p}) +  f_{i} ({\bf k},t)
\end{equation}

Where $M_{ijl}=k_{j}P_{il}({\bf k}) + k_{l}P_{ij}({\bf k})\/$
and the transverse projector, $P_{ij}= \delta_{ij} - 
k_{i}k_{j}/{k^2}\/$,
where the external noise $f_{i}\/$ is $\delta \/$ correlated and
is necessary to maintain the energy balance in the inertial 
range. The energy input per unit time ($\bar{\epsilon}\/$) at 
the long wave lengths cascades through different lenght scales 
due to the nonlinear term and for $k > k_{D}\/$, is dissipated by 
molecular viscosity ($\nu\/$), here $k_{D}= (\bar{\epsilon}
/\nu^{3})^{1/4}\/$. For ($k<<k_{D}\/$), we have the so called 
inertial range, where one expects,

\begin{equation}
S_{2n}\sim k^{-(\zeta_{2n} + 3n)}
\end{equation}

with structure function defined as in Dhar et al \cite{ourprl} as,
$S_{n}= \langle|v(k)|^{n}\rangle\/$. The exponent $\zeta_n\/$
is $n/3\/$ in the Kolmogorov limit. In general it differs from
$n/3\/$ and one of the best estimates of the deviation is due to 
She and Leveque \cite{shelev} which gives,

\begin{equation}
\delta\zeta_n= \zeta_{n}-n/3=-\frac{2n}{9}+2[1-(\frac{2}{3})^{n/3}]
\end{equation}

In this work we investigated the dissipation range and our
principal results are

\begin{equation}
S_{2n}(k)\sim k^{n\delta_{2}}e^{-nk/K},\;\;\;\;	(k>>K)
\end{equation}

where $\delta_{2} = 2-D\/$ (D being the dimensionality of space)
and $K= \Theta(k_{D})\/$. By studying the correction to the 
above result as powers of $K/k\/$, we propose (in $D=3\/$) the
crossover function (crossover from dissipation to inertial range)

\begin{equation}
S_{2}(k)\sim \frac{1}{k}(1+\alpha_{1}\frac{K}{k})^{2+\zeta_{2}}e^{-k/K}
\end{equation}

where $\alpha_{1}\/$ is number of $\Theta(1)\/$, while for higher
order structure function,

\begin{equation}
S_{2n}(k)\sim k^{n\delta_{2}}(1+\alpha_{n}\frac{K}{k})
^{ n(3+\delta_2)+ \zeta_{2n} }e^{-nk/K}
\end{equation}

The constants $\alpha_{n}\/$ are non-universal but will be shown
to be almost independent of $n\/$. The explicit crossover forms
that we have written down helps us understand the idea of extended
self similarity (ESS) introduced by Benzi et al\cite{benzi}. Our
approach is alternative to that of Segel et al\cite{lvpess}.
Writing Eq.6 by expanding about the inertial range form, we note
that,

\begin{equation}
S_{2n}(k)\sim k^{-(\zeta_{2n} + 3n)}
(1- [n - \frac{n(3+\delta_2)+ \zeta_{2n}}{\alpha_n}] \frac{k}{K})
\end{equation}

the simple power law will break down when,

\begin{eqnarray}
k\sim \frac{\alpha_n}{n(\alpha_n-3-\delta_2)-\zeta_{2n}}K \nonumber
\end{eqnarray}

From phenomenology, it is known that $S_n\/$  falls off from 
the $k^{-(\zeta_{2n} + 3n)}\/$ line in the dissipation range. 
This constrains $\alpha_n\/$ (from Eq 7 and using $\delta_{2} 
= 2-D\/$),

\begin{eqnarray} 
\frac{n(\alpha_n-2)-\zeta_{2n}}{\alpha_n} > 0\nonumber
\end{eqnarray}
 
Now if we assume $\alpha_n > 0 \/$ and use the fact that 
$\alpha_n\/$ is almost independent of $n\/$ (shown later), 
we get,

\begin{eqnarray} 
\alpha_1> 2 + \zeta_2 \nonumber
\end{eqnarray}

As $\alpha_n-2 > \zeta_2 (=2/3)\/$, the difference 
$n(\alpha_n-2)-\zeta_{2n}\/$ will grow with $n\/$ (since we
know that $\zeta_{2n}\/$ deviates more from linearity for
higher moments). Hence for higher $n\/$ the $S_n\/$ curves will 
fall off from the scaling regime at even lower $k\/$ values. 
This is completely consistent with the standard phenomenology 
\cite{Sn}.\\

We now turn to ESS. From Eq.6 it is clear that,

\begin{eqnarray}
S_{2n}(k)\sim &[&S_{2m}(k)]^{(\frac{\zeta_{2n}+3n}{\zeta_{2m}+3m})}
e^{-nk/K}e^{m(\frac{\zeta_{2n}+3n}{\zeta_{2m}+3m})k/K } \nonumber \\
& & \nonumber \\
& &\times\frac{ [1+\alpha_{n}^{-1}\frac{k}{K}]^{n(3+\delta_2)+ 
\zeta_{2n}}}
{[1+\alpha_{m}^{-1}\frac{k}{K}]^{(m(3+\delta_2)+ \zeta_{2m})
(\frac{\zeta_{2n}+3n}{\zeta_{2m}+3m})}}
\end{eqnarray}

The explicit $k\/$ dependant terms on the r.h.s. of the above 
expression will cause deviation from scaling. But it is apparent 
that the exponential factor is much more weakly decaying than 
$e^{-nk/K}\/$ (in fact it is constant for the Kolmogorov situation 
of $\zeta_{2n}=2n/3\/$) and also the variation of 
$[1+ \alpha_{n}^{-1}\frac{k}{K}]^{n(3+\delta_2)+ \zeta_{2n}}\/$ 
is muted by the denominator (as $\alpha_{n}\/$s' have been assumed 
to have the same sign and shown to be almost independent later). 
Consequently a plot of $\log S_{2n}\;\/$ 
vs $\;\log S_{2m}\/$ will remain a straight line over a far 
longer range than $S_{2n}\;\/$ vs $\;k^{-(\zeta_{2n} + 3n)}\/$. 
This is the content of ESS. Few other phenomenological consequences
are also manifest. For example, as $(n-m)\/$ grows the scaling 
regime will become gradually shorter. In fact with $\alpha_{n}\/$ 
independent of $n\/$ to a first approximation and $\zeta_{2n}\/$ 
almost proportional to $n\/$, the scaling of $\log S_{2n}\;\/$ 
vs $\;\log S_{2m}\/$ is virtually exact.\\

We first note that correlation functions in the dissipation 
range falls off extremely fast \cite{kraichnan} with the 
characteristic scale $k_{D}\/$ and because of the 
existance of the scale there is no divergence in the 
self energies and correlation functions. Absence of 
divergence in the self energy implies that vicosity
coefficient $\nu\/$ is not renormalised. The correlation 
function is given at the self consistent single loop level by,

\begin{eqnarray}
S_{2}(k,w)= |G|^{2}k^2 &\int& \frac{d^{D}p}{(2\pi)^D}
\frac{dw'}{2\pi} a({\bf k,p,k-p})\nonumber \\
 & & \times S_{2}(|{\bf k-p}|,w-w')S_{2}(p,w') 
\end{eqnarray}

where the angular factor, 
\begin{eqnarray}
a({\bf k,p,k-p})= \frac{1}{2}(1-xyz-2y^2z^2) \nonumber 
\end{eqnarray}

The trio $({\bf k,p,k-p})\/$ forms a triangle and $x,y,z\/$ 
are the direction cosines of the angles opposite to 
${\bf k,p}\/$ and ${\bf k-p}\/$ respectively. 
The response function $G^{-1}= -iw + \nu k^{2}\/$ and the 
correlation function $S_2(k,w)= k^{\delta_2} f(k/k_D)
k^2/(w^2 + \nu^2 k^4)\/$, such that

\begin{equation}
\int \frac{dw}{2\pi}S_{2}(k,w) = S_{2}(k,t=0)=
k^{\delta_2}f(k/k_{D})
\end{equation}

Comparing the two sides of Eq.9, the function $f(k/k_{D})\/$
has the structure $f(k/k_{D})= e^{-\beta k/k_{D}}\/$, since 
on the right hand side of Eq.9,
 
\begin{eqnarray}
e^{-\beta p/k_{D}} e^{-\beta |{\bf k-p}|/k_{D}} &=& 
e^{-\beta( {\bf |\frac{k}{2}-q| + |\frac{k}{2}+q|} )/k_{D}}
\nonumber \\
&=& e^{-\beta k/k_{D}} ( e^ {-\Theta(q^2/k^2)} + \ldots )
\nonumber \\ &\simeq& e^{-\beta k/k_{D}}
\end{eqnarray}

thus reproducing the exponential factor of the left hand side.
Power counting of the momentum in Eq.9 now leads to,
 
\begin{equation}
\delta_2 = -(D-2)
\end{equation}

To check the correctness of this self consistent solution 
we avaluated the equal time limit of the integral on the r.h.s. 
of eqn.9 numerically (with lower cutoff $k_D\/$ and $D=3\/$). 
In fig.1(a) we plot this integral $I_d(k)\/$ and 
compare it with the function $\frac{1}{k}e^{-k/k_D}\/$.  
The agreement is good for $k/{k_D}\geq 20\/$ 
(ie, $k>>{k_D}\/$).\\


If the above formalism has to approach the crossover
behaviour, then we need to include the first correction to
the large $k\/$ behaviour. We do this by saying that the 
correction is in powers of $(K/k)^m\/$ and thus in $D=3\/$

\begin{equation}
S_2 = \frac{b_0}{k} [ 1 + b_1(K/k)^m] e^{-k/K}
\end{equation}

where $K=k_D/\beta\/$. The right hand side of Eq.9, 
linearised in $b_1\/$ and considered at zero frequency can be 
written as being proportional to:

\begin{eqnarray}
&\int & \frac{d^{3}p}{(2\pi)^3} a({\bf k,p,k-p}) 
\frac{ e^{-(p+ |{\bf k-p}|)/K} }{ (p^2 + |{\bf k-p}|^2) 
p |{\bf k-p}| } \nonumber \\
& &\times [ 1 + b_1(K/p)^m + b_1(K/|{\bf k-p}|)^m ]
\end{eqnarray}

\begin{figure*}[t]
\centerline{\epsfxsize=19pc \epsfbox{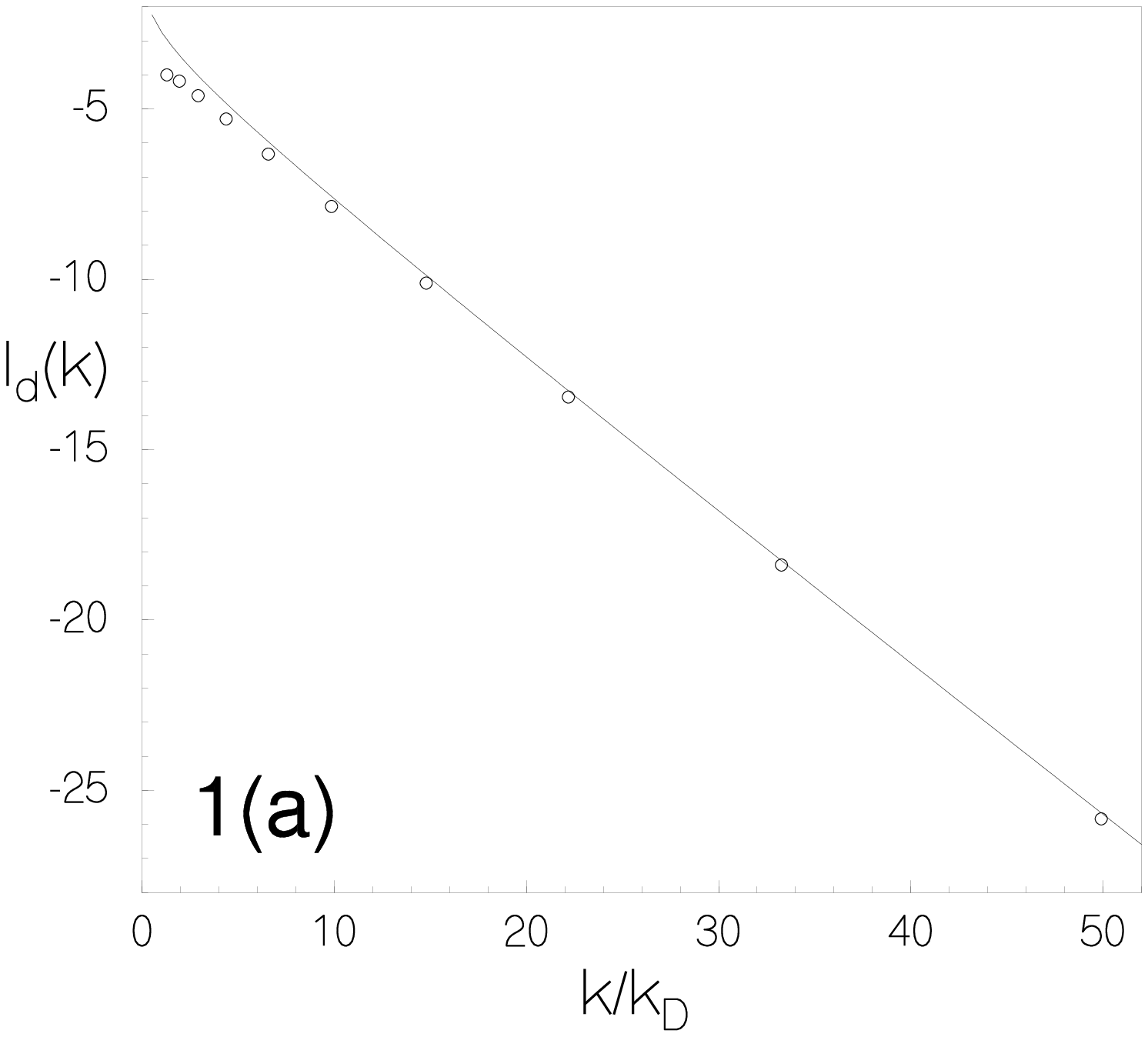} \hfill
\epsfxsize=19pc \epsfbox{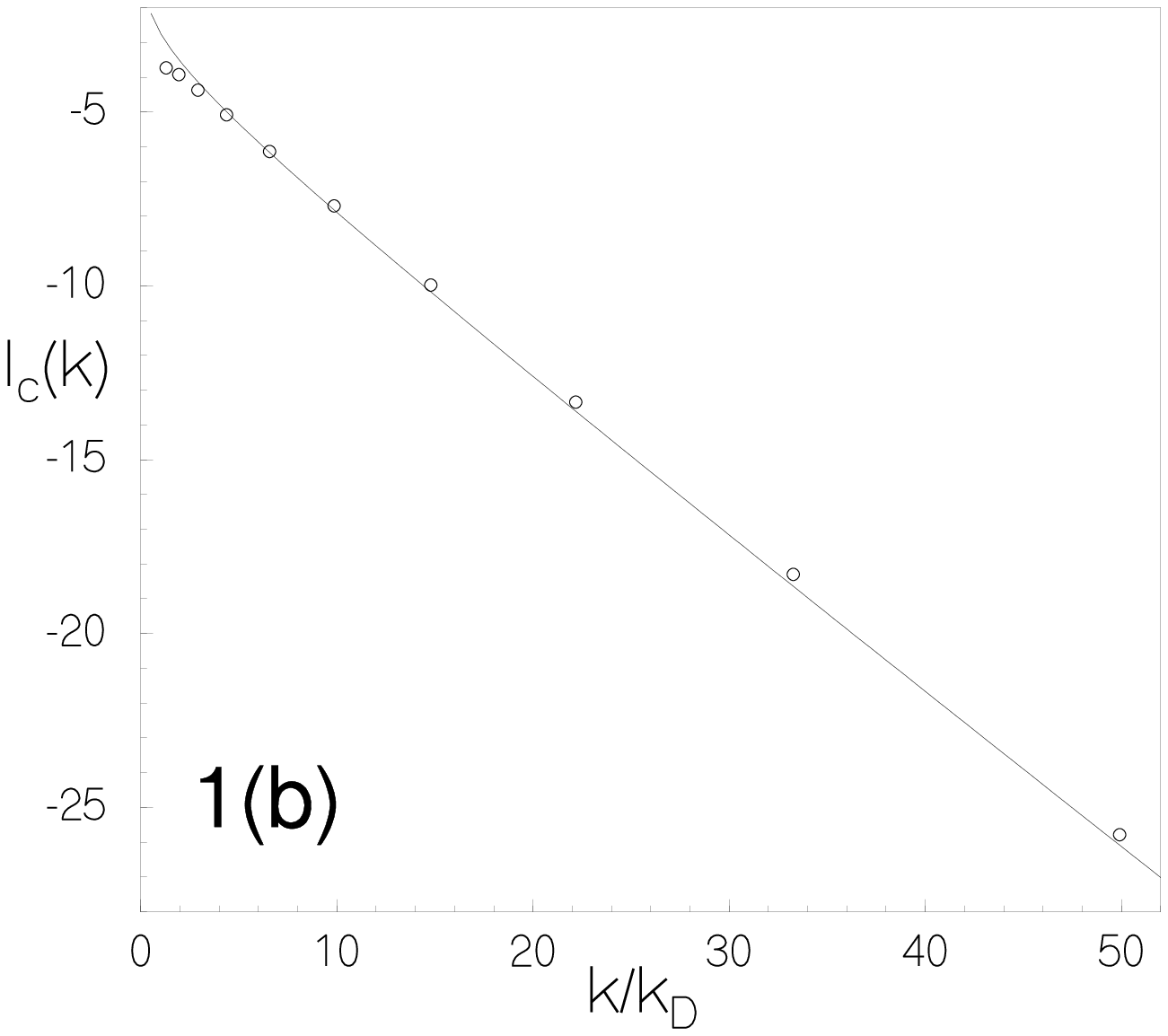}}
\caption{The dominant $\Theta ({b_1}^0)\/$ and correction
$\Theta (b_1)\/$ terms in eq.14, obtained from numerical 
integration (circle 'o') are compared with the respective 
terms (solid line) in eqn.13. 
(a)circle: $log_{10} I_d(k)\/$ ; 
solid: $\frac{b_0}{k} e^{-k/k_D}\/$
(b) $log_{10} I_c(k)\/$ ; 
solid: $\frac{b_0}{k}({k_D}/k)^{1/4} e^{-k/k_D}\/$.}
\label{fig1}
\end{figure*}

The requirement that the integral involving $b_1\/$ is finite,
leads to $m<2\/$. If we write $m=2-\epsilon\/$, we can 
evaluate the integrals to the leading pole \cite{jkbferl} 
in $\epsilon\/$. The integral not involving $b_1\/$ can be
evaluated using saddle point technique, with dominant 
contribution coming from $p\simeq k\/$. The result of the 
above manipulation must 
be of the form $[1+ b_1(K/p)^m]\/$ for self consistency and 
at the level of approximation just described, we find 
$m=1\/$. Thus, we have the result that for $k>>k_D,\;\;  
S_2\propto \frac{1}{k}(1+b_1\frac{K}{k})e^{-k/K}\/$ and for
$k<<k_D\/$ (inertial range), $S_2\propto k^{-(3+\zeta_2)}\/$.
The simplest interpolation is Eq.5.\\

Using the above analysis as a guide towards determining 
$m\/$, we numerically evaluated the correction integral 
$I_c(k)\/$ of $\Theta (b_1)\/$ in eqn.14. Using the same 
values of $b_0\/$ and the lower cutoff $k_D\/$ which we had 
used for fitting the dominant term, we find self 
consistency of the correction integral can 
be achieved for $m \simeq 1/4\/$. In fig.1(b) we plot this 
integral $I_c(k)\/$ as a function of $k/{k_D}\/$ and compare it 
with the $\Theta (b_1)\/$ term in eqn.13. The agreement is good
for $k/{k_D} \geq 7\/$. We have chosen $K = {k_D}\/$ for our 
numerics. All our arguments demonstrating the ESS properties 
of $S_n\/$ hold good as long as $m > 0\/$. 

 It should be noted that in this far dissipation range that
we are considering here, the single loop self consistency 
is sufficient. We have checked that the contributions from 
higher ($\geq 2\/$) loop diagrams are at most of the same 
order as the single loop diagram. 
So their inclusion just changes the amplitude of $S_{2}(k)\/$. 
This statement is true for the evaluation of $S_{2n}\/$  
($n>1\/$) also, which we do now. Out of the various possible 
arrangements of the ${\bf k}\/$ and ${\bf -k}\/$ external 
legs on an one loop diagram, we evaluate the most relevant 
one (shown in Fig.1). Contribution from other possible one 
loop diagrams are exponentially smaller and hence their 
contributions are negligible. The contribution from Fig.1 is,

\begin{eqnarray}
S_{2n}(k,t)&=&\langle [v({\bf k},t)v({\bf -k},t)]^n\rangle
\nonumber \\ &\sim& k^{2n}\int_{-\infty}^{t}dt_1\ldots
\int_{-\infty}^{t}dt_{2n}\int \frac{d^{D}p}{(2\pi)^D}\nonumber \\
& &\times G(k;t,t_1)G(-k;t,t_2)\ldots \nonumber \\
& &\times S_2(p,|t_1-t_2|)S_2(|{\bf k-p}|,|t_2-t_3|)\ldots
\end{eqnarray}

As $S_2(k,t)\sim k^{-(D-2)}e^{-k/K}e^{-|t|\nu k^2}\/$, 
we note that the integral of Eq.15 will be 
dominated by the low momentum pole at $p=k\/$. Using a pole
approximation for evaluating the integral, a momentum count
produces the result that $S_{2n}(k,t)\propto k^{n\delta_{2}}
e^{-nk/K}\/$ . This establishes Eq 4. However, within this
formalism, though we cannot rigorously show that Eq.4 holds for 
odd moments also, for monotonicity sake we assume this to be 
true. Now we note that Eq 4 implies $S_n \sim S_{3}^{n/3}\/$ ie, 
simple scaling behaviour results in the far dissipation range.
This is in mild contrast to the simulation results \cite{ourprl},
where very weak multiscaling (ie, very small deviation from
$n/3\/$) has been reported.  But given that this deviation 
is very small, e.g. for $n=7\/$ the numerical exponent is 2.24 
instead of 7/3, our estimate for this far dissipation 
range is a very close one.\\

\begin{figure}[t]
\epsfxsize=19pc
\centerline{\epsfbox{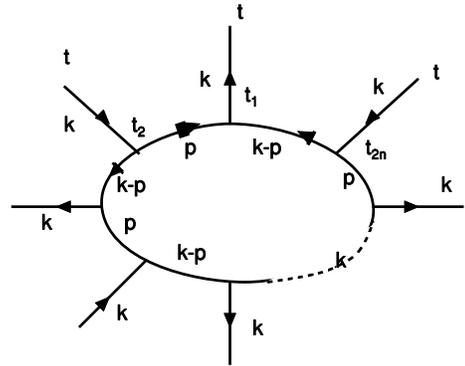}}
\caption{Diagram contributing to $S_{2n}(k)\/$ in the 
dissipation regime}
\label{fig2}
\end{figure}

We now study the first deviation of $S_{2n}\/$ from its form
in Eq 4. To do so, we introduce the first deviation of
$S_{2n}(k,t)\/$ in Eq 15. The integral in in Eq.15 is 
already pole dominated and hence the additional part is
pole dominated as well. There are $n\/$ contributions
of equal strength from each of the $S_2(p)\/$ and 
$S_2(|{\bf k-p}|)\/$  and consequently for $k>>K\/$ 

\begin{equation}
S_{2n}\propto k^{-n(D-2)} [ 1 + b_n(K/k)] e^{-nk/K}
\end{equation}

where $b_n \propto nb_1\/$. With the  quantity 
$n(3+\delta_2) + \zeta_{2n}\/$ in Eq.6 roughly proprtional to 
$n\/$, we consequently infer that in the interpolation 
formula of Eq.6, the constant $\alpha_n\/$ is to a good 
approximation independent of $n\/$. Thus the main results 
Eq.4 - Eq6, are obtained.\\

Now we look at the real space structure function $S_2(r)=
\langle[{\bf v}({\bf x}+{\bf r})-{\bf v}({\bf x})]^2 
\rangle\/$ which is the inverse Fourier transform of 
 $2[{u_0}^2 \delta(k) - S_2(k)]\/$ (where ${{u_0}^2}/2\/$ is 
the mean energy). For $r\/$ in the far dissipation range 
$S_2(r)\/$ will be determined by our $k>>k_d\/$ form of 
$S_2(k)\/$ (ie,$\sim \frac{1}{k}e^{-k/K}\/$).
This yields $S_2(r) = c_1 r^2 + \Theta (r^4)\/$. Here
$c_1\/$ is a function of $\nu,\bar{\epsilon}\/$.
This form of $S_2(r)\/$ is consistent with the result of 
Sirovich et.al.\cite{sirov}. The added advantage of 
our $k -\/$ space calculation is the ability to predict 
the higher order structure functions ($S_{2n}(k)\/$) also.\\

In summary we have shown that by considering Navier Stokes
equation and doing a self consistent treatment of the 
dissipation range (characterised by the existence of a
scale $k_D\/$ ), we can establish forms for the various
order structure functions. By the first correction to the
asymtotic situation and using the known results in the 
inertial range ($k<<k_D\/$), we can construct explicit
crossover functions for the structure functions (crossover
from $k>>k_D\/$ to $k<<k_D\/$ ). The validity of ESS is 
easy to see.\\

We would like to thank R.Pandit for comments, P.Pradhan 
for help in drawing the Feynman diagram (fig.2) 
and CSIR (India) for support.

\end{document}